\documentclass{article}

\usepackage{arxiv}
\usepackage[utf8]{inputenc} % allow utf-8 input
\usepackage[T1]{fontenc}    % use 8-bit T1 fonts

\usepackage{hyperref}       % hyperlinks
\hypersetup{
  colorlinks   = true, %Colours links instead of ugly boxes
  urlcolor     = blue, %Colour for external hyperlinks
  linkcolor    = blue, %Colour of internal links
  citecolor    = blue  %Colour of citations
}
\usepackage{url}            % simple URL typesetting
\usepackage{cite}
\usepackage{amsmath}
\usepackage{amsthm}         % provide math theorems
\usepackage{amssymb}        % provide math symbols such as \triangleq
\usepackage{booktabs}       % professional-quality tables
\usepackage{amsfonts}       % blackboard math symbols
\usepackage{nicefrac}       % compact symbols for 1/2, etc.
\usepackage[activate={true,nocompatibility},
            final,
            tracking=true,
            kerning=true,
            spacing=true,
            factor=1100%,
            ]{microtype}
\usepackage{xpunctuate}     % macros with optional periods, commas, etc
\usepackage{caption}        % customize caption typesetting

\usepackage{algorithm}
\usepackage{algpseudocode}
\usepackage{ifthen}         % conditionals
\usepackage{graphicx}
\usepackage{doi}
\usepackage[right]{lineno}  % line numbers
\usepackage{cancel}

\providecommand{\eg}[0]{\xperiodcommaafter{\textit{e.g}}}

\providecommand{\etal}[0]{\xperiodafter{\textit{et~al}}}

\providecommand{\norm}[1]{ {\left\lVert #1 \right\rVert} }
\providecommand{\abs}[1]{\left| #1 \right|}
\providecommand{\prth}[1]{\left( #1 \right)}
\providecommand{\brcs}[1]{\left\{ #1 \right\}}
\providecommand{\sqbr}[1]{\left[ #1 \right]}

\providecommand{\equn}[0]{Eq\xperiod}

\providecommand{\figr}[0]{Fig\xperiod}

\providecommand{\secn}[0]{Sec\xperiod}

\providecommand{\code}{\texttt}

\algrenewcommand{\Return}{\State\algorithmicreturn~}
\algnewcommand\algorithmicinput{\textbf{Input:}}
\algnewcommand\Input{\item[\algorithmicinput]}
\algnewcommand\algorithmicoutput{\textbf{Output:}}
\algnewcommand\Output{\item[\algorithmicoutput]}
\algnewcommand\algorithmicinitialize{\textbf{Initialize:}}
\algnewcommand\Initialize{\item[\algorithmicinitialize]}

\newcommand{\ifempty}[3]{%
    \ifthenelse{ \equal {#1} {} }
        {#2} % if #1 is empty
        {#3} % else (not empty)
    }

\newcommand{\subscriptmathifnotempty}[2]{%
    \ifempty {#1}
        {\ensuremath{#2}}
        {\ensuremath{{#2}_{#1}}}
}

\newtheorem*{plrtheorem}{PLR Theorem}

\DeclareFontFamily{U}{mathx}{\hyphenchar\font45}
\DeclareFontShape{U}{mathx}{m}{n}{<-> mathx10}{}
\DeclareSymbolFont{mathx}{U}{mathx}{m}{n}
\DeclareMathAccent{\widebar}{0}{mathx}{"73}

\providecommand{\transpose}{\ensuremath{^{\mathsf{T}}}}
\providecommand{\identity}{\mathbb{I}}
\providecommand{\realnum}{\mathbb{R}}

\DeclareMathOperator*{\argmax}{arg\,max}
\DeclareMathOperator*{\argmin}{arg\,min}
\providecommand{\expectation}[2][]{%
    \ifthenelse { \equal {#1} {} }
        {\mathbb{E}\kern-0.1ex\sqbr{#2}}      % if #1 == blank
        {\mathbb{E}_{#1}\kern-0.1ex\sqbr{#2}} % else (not blank)
    }
\providecommand{\prob}[1]{p\kern-0.2ex\prth{\, #1\,}}
\providecommand{\posterior}[1][]{%
    \ifthenelse { \equal {#1} {} }
        {\ensuremath{p_{\textrm{post}}}}
        {\ensuremath{p_{\textrm{post}}\kern-0.2ex\prth{\, #1 \,}}}
    }

\providecommand{\latvect}[1]{\mathbf{#1}}
\providecommand{\symvect}[1]{\boldsymbol #1}

\providecommand{\vecb}[0]{\latvect{b}}
\providecommand{\vecd}[0]{\latvect{d}}
\providecommand{\vecf}[0]{\latvect{f}}

\providecommand{\vecH}[0]{\latvect{H}}
\providecommand{\vech}[0]{\latvect{h}}
\providecommand{\vecr}[0]{\latvect{r}}

\providecommand{\vecv}[0]{\latvect{v}}

\providecommand{\vecx}[0]{\latvect{x}}
\providecommand{\vecX}[0]{\latvect{X}}
\providecommand{\vecy}[0]{\latvect{y}}
\providecommand{\vecz}[0]{\latvect{z}}
\providecommand{\G}[0]{\mathcal{G}}

\providecommand{\vecmu}[0]{\symvect{\mu}}
\providecommand{\vectheta}[0]{\symvect{\theta}}
\providecommand{\vecSigma}[0]{\symvect{\Sigma}}

\providecommand{\KL}[2]{D_{\mathrm{KL}}\prth{ #1 \parallel #2 }}

\DeclareMathOperator{\NM}{NM}
\providecommand{\NormalMeans}[2][]{%
    \ifthenelse { \equal {#1} {} }
        {\NM\kern-0.2ex\prth{#2}}
        {\NM_{#1}\kern-0.2ex\prth{#2}}
    }
\providecommand{\mllNM}[0]{\ensuremath{\ell_{\NM}}}
\providecommand{\elboNM}[1][]{%
    \ifempty {#1}
        {\ensuremath{ F^{\NM} }}
        {\ensuremath{ F^{\NM}\kern-0.2ex\prth{\,#1\,} }}
    }

\providecommand{\shrinkageop}[1][]{\subscriptmathifnotempty{#1}{S}}
\providecommand{\shrinkageopinv}[1][]{\subscriptmathifnotempty{#1}{T}}
\providecommand{\penaltyop}[1][]{\subscriptmathifnotempty{#1}{\rho}}
\providecommand{\vbbar}[0]{\vectheta}
\providecommand{\bjbar}[0]{\theta_j}
\providecommand{\htilde}[0]{\widetilde{h}}

\begin{document}

% ========================================================== 
% Title requires different formatting for different journals
% and hence included in journal-specific directories
% \input{arxiv/title}
\title{Gradient-based optimization for variational empirical Bayes multiple regression}

\author{\href{https://orcid.org/0000-0003-4437-8833}%
            {\includegraphics[scale=0.06]{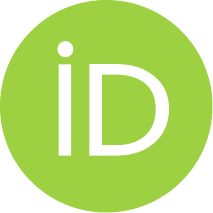}}\hspace{2mm}%
            Saikat Banerjee
            %\thanks{Use footnote for providing further information about author (webpage, alternative address)}%
            \\
            Department of Human Genetics\\
            University of Chicago\\
            \texttt{saikat.banerjee@uchicago.edu} \\
       \And
       \href{https://orcid.org/0000-0003-1144-6780}
            {\includegraphics[scale=0.06]{orcid.pdf}}\hspace{2mm}%
            Peter Carbonetto
            \\
            Department of Human Genetics, and\\
            Research Computing Center\\
            University of Chicago\\
            %Chicago, IL 60637 \\
            \texttt{peter.carbonetto@uchicago.edu} \\
       \And
       \href{https://orcid.org/0000-0001-5397-9257}
            {\includegraphics[scale=0.06]{orcid.pdf}}\hspace{2mm}%
            Matthew Stephens
            \\
            Department of Human Genetics, and\\
            Department of Statistics\\
            University of Chicago\\
            \texttt{mstephens@uchicago.edu}\\
}
\renewcommand{\undertitle}{}
\date{}

% Header on other pages
\renewcommand{\headeright}{}
\renewcommand{\shorttitle}{GradVI}

\maketitle
% ==========================================================

\begin{abstract}
\noindent
Variational empirical Bayes (VEB) methods provide a practically 
attractive approach to fitting large, sparse, multiple regression models.
These methods usually use coordinate ascent to optimize the variational objective function,
an approach known as coordinate ascent variational inference (CAVI). 
Here we propose alternative optimization approaches based on gradient-based (quasi-Newton) methods, 
which we call gradient-based variational inference (GradVI).
GradVI exploits a recent result from Kim \etal [arXiv:2208.10910] 
which writes the VEB regression objective function as a penalized regression. 
Unfortunately the penalty function is not available in closed form, 
and we present and compare two approaches to dealing with this problem.
In simple situations where CAVI performs well, 
we show that GradVI produces similar predictive performance,
and GradVI converges in fewer iterations when the predictors are highly correlated.
Furthermore, unlike CAVI, the key computations in GradVI are simple matrix-vector products, 
and so GradVI is much faster than CAVI in settings where the design matrix 
admits fast matrix-vector products (\eg, as we show here, trendfiltering applications)
and lends itself to parallelized implementations in ways that CAVI does not. 
GradVI is also very flexible, and could exploit automatic differentiation 
to easily implement different prior families.
Our methods are implemented in an open-source Python software, GradVI
available from \href{https://github.com/stephenslab/gradvi}{https://github.com/stephenslab/gradvi}.
\end{abstract}

% ==========================================================
% Include line numbers for preprints
%\linenumbers
% ========================================================== 

% ##########################################################
% \input{main/introduction.tex}
\section{Introduction}
% ##########################################################

Multiple linear regression provides a simple, but widely used, method
to find associations between outcomes (responses) and a set of predictors (explanatory variables).
It has been actively studied over more than a century, and
there is a rich and vast literature on the subject~\cite{stigler_1984}.
In practical situations the number of predictor variables is often large, and it becomes desirable to induce sparsity in the regression coefficients to avoid overfitting~\cite{titterington_high_dimensional_data_2009, hastie_2015_statistical_learning_with_sparsity}.
Sparse linear regression also serves as the foundation for non-linear techniques, such as trendfiltering \cite{kim_2009_siamrev_l1_trend_filtering, tibshirani_2014_annstat_trend_filtering},
which can estimate an underlying non-linear trend from time series data.
Applications of sparse multiple linear regression and trendfiltering arise in a wide range of applications in modern science and engineering, including astronomy~\cite{feigelson_babu_2012_astronomy_statmeth}, atmospheric sciences~\cite{von_storch_2002_climate_statmeth}, biology~\cite{buehlmann_2014_annrev_stat_bio}, economics~\cite{tsay_2005_financial_time_series, primeceri_2021_economics_big_data}, genetics~\cite{guan_stephens_2011, carbonetto2012-bvsr, wang_2020_susie, zhou_2013_bslmm, qian_2020_plogen_basil}, geophysics~\cite{gubbins_2004_geophysics_time_series}, medical sciences~\cite{cawley_talbot_2006_cancer_sparse_logistic_regression, zeger_2006_annrev_biomedical_time_series}, social sciences~\cite{hindman_2015_social_science_better_models} and text analysis~\cite{taha_2024_text_regression_review}.

Approaches to sparse linear regression can be broadly classified into two groups:
(a) penalized linear regressions (PLR), which add a penalty term to the likelihood
to penalize the magnitude of its parameters~\cite{hoerl_1970_ridge,
tibshirani_1996_jrss_lasso, zou_hastie_2005_elastic_net}, and
(b) Bayesian approaches~\cite{mitchell_1988_bayesian_variable_selection, mcculloch_1993_variable_selection_gibbs_sampling,
park_casella_2008_bayesian_lasso, li_lin_2010_bayesian_elastic_net, guan_stephens_2011, carbonetto2012-bvsr, zhou_2013_bslmm, stephens_2016_fdr_new_deal, wang_2020_susie, kim_mrash_2022}, which use a prior probability distribution on the model parameters to induce sparsity. Computation for PLR approaches naturally involves optimization algorithms, including both coordinate ascent \cite{friedman_2010_lasso_regularization} and gradient-based methods \cite{leCessie_houwelingen_1992_newton_raphson_ridge, frank_wolfe_1956, tong_2004_icml_stochastic_gradient_descent}. Computation for Bayesian sparse regression is usually done either via Markov chain Monte Carlo \cite{mcculloch_1993_variable_selection_gibbs_sampling, guan_stephens_2011} or by variational approximation \cite{carbonetto2012-bvsr}.

Bayesian and PLR approaches are usually seen as distinct classes of method. Although some PLR methods may be interpreted as {\it maximum a posteriori} (MAP) estimates under some prior (eg lasso is MAP estimation under a Laplace prior \cite{park_casella_2008_bayesian_lasso}), most Bayesian approaches use sparse priors with a point mass at 0, and MAP estimation does not work with such priors (the MAP is always 0 because the density at 0 is infinite). Thus Bayesian approaches to sparse regression do not appear to correspond to a PLR. However, in recent work
Kim \etal \cite{kim_mrash_2022} showed how variational approximation methods for Bayesian sparse linear regression can indeed be interpreted as a PLR.
Since variational approximation methods frame the Bayesian inference problem as an optimization problem (they seek an optimum approximation to the posterior distribution in some approximating family) it is perhaps not surprising that they are connected with PLRs. However, variational methods optimize over an approximation to the posterior distribution, whereas PLR methods optimize over the regression parameters, so the two approaches appear different. Kim \etal connected the two approaches by writing the variational optimization problem as an optimization over the posterior mean of the regression parameters, and showing how this could be interpreted as a PLR.

Kim \etal's variational method also used an Empirical Bayes (EB) approach to estimate the prior distribution on the regression parameters from the data. They point out that the resulting variational Empirical Bayes (VEB) approach can be seen as a PLR in which the penalty, which depends on the prior, is tuned by directly solving an optimization approach. This approach has the advantage that it can tune very flexible priors/penalties with many tuning parameters; in contrast, the conventional PLR approach of tuning penalties via cross-validation is usually restricted to a single tuning parameter. \cite{kim_mrash_2022} show that by using a flexible family of prior distributions -- a finite mixture of normal distributions, referred to as the ``adaptive shrinkage'' (ash) prior in \cite{stephens_2016_fdr_new_deal} -- the VEB approach
can provide competitive performance across a wide range of sparsity levels compared with commonly-used PLR methods such as lasso~\cite{tibshirani_1996_jrss_lasso}, ridge regression~\cite{hoerl_1970_ridge} and elastic net~\cite{zou_hastie_2005_elastic_net}.

\begin{figure}[t]
  \includegraphics[width=\textwidth]{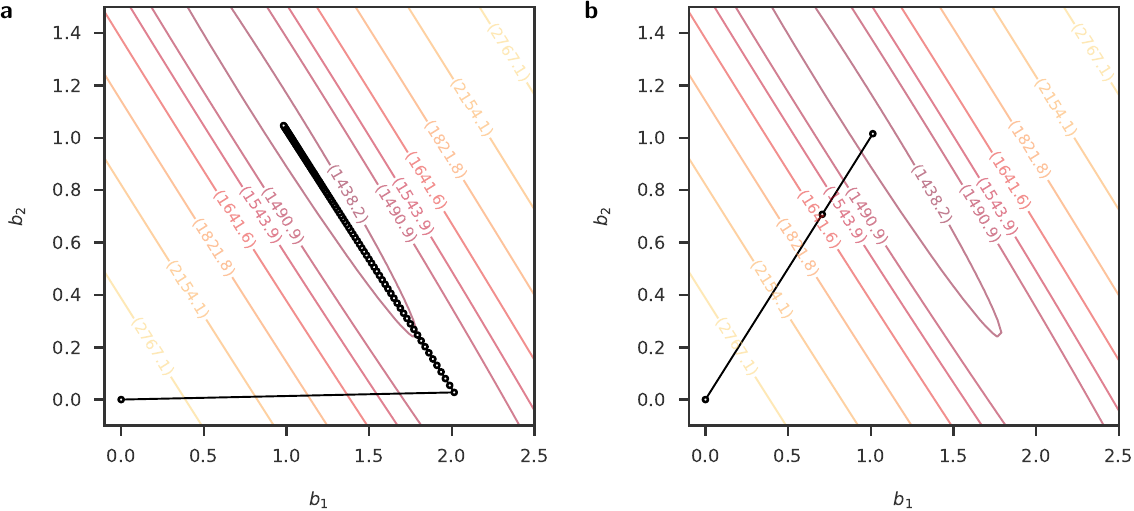}
  \caption{
    \textbf{Optimization steps in a toy example with two correlated predictors.}
    We simulated 1000 samples from a multiple linear model $\vecy = b_1 \vecx_1 + b_2 \vecx_2 + \symvect{\epsilon}$
    where $\vecx_1$ and $\vecx_2$ are strongly correlated (Pearson correlation coefficient $\rho = 0.98$),
    the regression coefficients are $b_1 = b_2 = 1.0$
    and the noise $\symvect{\epsilon}$ is sampled from a standard normal distribution.
    In the plots, we show the progress of optimization to estimate $b_1$ and $b_2$ using
    (a) coordinate ascent (left panel) and (b) quasi-Newton (right panel) methods,
    given a fixed prior and residual variance.
    The contours show the negative ELBO as a function of $b_1$ and $b_2$ (on the $x$ and $y$ axis respectively).
    The black circles show the estimated values of $\widehat{b}_1$ and $\widehat{b}_2$ in each step of the algorithms,
    the black line connects consecutive steps.
    }
  \label{fig:2d-optim-steps}
\end{figure}

Following the most commonly-used approach in variational inference, \cite{kim_mrash_2022} used coordinate ascent to optimize the variational objective function. The resulting ``coordinate ascent variational inference" (CAVI) algorithm  performed well in their numerical comparisons. However, CAVI has some important computational limitations. First, CAVI can be slow to converge in settings with correlated predictors (Figure \ref{fig:2d-optim-steps}).  Second, because the CAVI algorithm, by design, updates one parameter at a time, holding all others fixed, it cannot be easily parallelized. It also cannot exploit opportunities for fast matrix-vector multiplication which can arise either due to special structure of the problem, as in our trend filtering examples later,
or due to use of specialized routines on GPUs.
These limitations may hinder the application of CAVI to very large
problems (\eg millions of predictors).

Motivated by these limitations of CAVI we introduce ``Gradient-based variational inference" (GradVI), which uses gradient-based optimization algorithms~\cite{nocedal_wright_2006} -- specifically,
quasi-Newton methods~\cite{davidon_technical_report_1959, davidon_1991_quasi_newton} -- to fit the VEB/PLR approach of \cite{kim_mrash_2022}.  GradVI retains similar predictive accuracy as CAVI, but is structured so that the
 main computational burden is matrix-vector products.
Thus, GradVI can immediately exploit any opportunities for fast matrix-vector multiplication,
making it considerably faster than CAVI in some settings. The gradient-based approach can, in principle, converge in much fewer iterations than CAVI in settings with correlated variables (Figure \ref{fig:2d-optim-steps}). The GradVI approach also has some other potential advantages: for example one could exploit automatic differentiation to easily implement different priors.

The remainder of the paper is organized as follows.
\secn \ref{sec:background} summarizes necessary background on the VEB method
of \cite{kim_mrash_2022} and the connection with PLR. Section
\secn~\ref{sec:method} describes
our methods for quasi-Newton optimization of the PLR objective, and \secn~\ref{sec:experiments} demonstrates
its efficacy using numerical experiments.
Finally, \secn~\ref{sec:discussion} summarizes the advantages and limitations of GradVI,
along with recommendations for its use.

% ##########################################################
% \input{main/methods.tex}
% ##########################################################

% ================================
% =================================
\section{Background: Coordinate Ascent Variational Empirical Bayes}
\label{sec:background}
% =================================
% =================================

\subsection{Notations}
We denote matrices by bold uppercase letters, \eg $\vecX$,
vectors by bold lowercase letters, \eg $\vecb$,
and scalars by letters, \eg $y$ or $Y$.
We use $\vecx_j$ to denote the $j$-th column of matrix $\vecX$.
We write the matrix transpose of $\vecX$ as $\vecX\transpose$,
the matrix inverse as $\vecX^{-1}$ and the matrix determinant as $\det(\vecX)$.
We write sets and families in calligraphic font, \eg $\mathcal{G}$.
We use the same convention for functions, \eg a scalar function of $\vecx$
is written as $f(\vecx)$ or $F(\vecx)$ while a vector function is written as $\vecf(\vecx)$.
We use $\realnum$ to denote the set of real numbers, $\realnum_+$ to denote its restriction to non-negative reals, and
$\identity_n$ to denote the $n \times n$ identity matrix. We use $\triangleq$ to indicate definitions. We use
$\mathcal{N}(
\mathbf{\mu},\mathbf{\Sigma})$ to denote the normal distribution with mean $\mathbf{\mu}$ and covariance matrix $\mathbf{\Sigma}$, and $\mathcal{N}( \cdot \mid
\mathbf{\mu},\mathbf{\Sigma})$ to denote its density. We use $\partial_{i} f$
to denote the partial derivative of the function $f$ with respect to its $i^{\textrm{th}}$ argument
\footnote{So, for example, given a function $h(x,y) \triangleq f(g(x, y), y)$, the partial derivative of $h$ with respect to $y$, evaluated at $(x,y)$, is written as
$\partial_2 h(x,y)$ and is equal to $\partial_1 f(g(x,y),y)\,\partial_2 g(x,y) + \partial_2 f(g(x,y),y)$.},
and we use $f'$ as shorthand for $\partial_1f$ (ie $f'$ indicates the partial derivative of $f$ with respect to its first argument).

% =================================
\subsection{Multiple Linear Regression}
% =================================
We consider a multiple linear regression model,
\begin{equation}
    \vecy \mid \vecX, \vecb, \sigma^2 \sim \mathcal{N}(\vecX\vecb, \sigma^2 \identity_n),
\label{eq:linear-model}
\end{equation}
where $\vecy \in \realnum^n$ is a vector of responses;
$\vecX \in \realnum^{n \times p}$ is a design matrix whose columns contain
predictors $\vecx_1, \ldots, \vecx_p \in \realnum^n$;
$\vecb \in \realnum^p$ is a vector of regression coefficients;
$\sigma^2 \geq 0$ is the variance of the residual errors.
While an intercept is not explicitly included in \eqref{eq:linear-model},
it is easily accounted for by centering $\vecy$ and the columns of $\vecX$ prior to model fitting.
We assume the regression coefficients $b_j$ are independent
and identically distributed (\textit{i.i.d.}) from some prior
distribution with density $g$, which is assumed to be a member of some prespecified family of distributions $\G$,
\begin{equation}
    b_j \sim g \in \G.
\label{eq:prior-general}
\end{equation}

% =================================
\subsection{Variational Empirical Bayes (VEB)}
% =================================

Given the prior $g$ and residual variance $\sigma^2$, Bayesian inference for the regression parameters $\vecb$ is based on the posterior distribution
\begin{equation}
    \posterior[\vecb \mid \vecy, \vecX, g,\sigma^2]
        \propto \prob{\vecy \mid \vecX, \vecb, \sigma^2}
                \prob{\vecb \mid g, \sigma^2}.
\label{eq:posterior_calculation}
\end{equation}
Unfortunately this posterior distribution is intractable for most priors of interest, and so must be approximated. Given some suitably-specified family $\mathcal{Q}$ of approximating distributions, variational methods seek to find the $q\in \mathcal{Q}$ that minimizes the Kullback–Leibler (KL) divergence from $q$ to the true posterior~\cite{carbonetto2012-bvsr, bishop-pattern-recognition-machine-learning}. That is, they seek
\begin{align}
    \widehat{q} &\triangleq \argmin_{q \in \mathcal{Q}}  \KL{q}{\posterior}
    \label{eq:variational-inference}
\end{align}
where $\KL{q}{p}$ denotes the KL divergence~\cite{kullback1951}
from a distribution $q$ to a distribution $p$.
Since $\KL{q}{\posterior}$ is itself computationally intractable, it is standard to recast the optimization problem \eqref{eq:variational-inference} as the equivalent maximization problem
\cite{jordan1999-intro-vi, bishop-pattern-recognition-machine-learning, blei2017-vi-review},
\begin{equation}
    \widehat{q} = \argmax_{q \in \mathcal{Q}}  F(q, g, \sigma^2)\,,
    \label{eq:variational-inference-optimization}
\end{equation}
where
\begin{align}
    F(q, g, \sigma^2) &= \log \prob{ \vecy \mid \vecX, g, \sigma^2 } - \KL{q}{\posterior}
    \label{eq:elbo-def-postKL} \\
              &= \expectation[q]{\log \prob{ \vecy \mid \vecX, \vecb, g, \sigma^2 }} - \KL{q}{g}.
    \label{eq:elbo-def-priorKL}
\end{align}
The function $F$ is often called the ``Evidence Lower Bound'' (ELBO) because it is a lower-bound for the ``evidence'',
$\log p\prth{\vecy \mid \vecX, g, \sigma^2}$.

The variational Empirical Bayes (VEB) approach builds on the above by also estimating $g, \sigma^2$ by optimizing $F$. That is, it optimizes
\begin{equation} \label{eq:veb}
    \widehat{q},\widehat{g},\widehat{\sigma}^2 \triangleq \argmax_{q \in \mathcal{Q}, g \in \mathcal{G}, \sigma^2 \in \realnum_+}  F(q, g, \sigma^2)\,.
\end{equation}

Kim \etal~\cite{kim_mrash_2022} used this VEB approach
to perform approximate Bayesian inference for multiple linear regression
with the adaptive shrinkage (normal mixture) prior~\cite{stephens_2016_fdr_new_deal}.
They use the so-called ``mean field" approximation \cite{carbonetto2012-bvsr}, which assumes the distributions in $\mathcal{Q}$ factorize as
\begin{equation}
    \mathcal{Q} = \brcs{q: q\prth{\vecb} = \prod_{j = 1}^{P} q_j(b_j)},
\end{equation}
and they develop a coordinate ascent algorithm for solving the resulting optimization problem
\eqref{eq:veb}. This algorithm (CAVI, for coordinate ascent variational inference) iterates through updating the variational factors $q_1,\dots,q_p$ until the ELBO $F(q, g, \sigma^2)$ converges; see Algorithm~\ref{alg:cavi-veb}.

\begin{algorithm}[htbp]
\caption{Coordinate ascent algorithm for VEB}
\label{alg:cavi-veb}
\begin{algorithmic}[0]
\Require {Data $\vecX \in \realnum^{n \times p}$, $\vecy \in \realnum^n$}
\Initialize {Variational factors $q_j(b_j)$, prior density $g$, error variance $\sigma^2$}
\Repeat
    \For {$j \in \brcs{1, \ldots, p}$}
        \State Set $q_j(b_j) \propto \exp\brcs{\expectation[q(\vecb_{-j})]{\log \prob{\vecy, \vecb \mid \vecX, g, \sigma^2}}}$
    \EndFor
    \State $g \gets \argmax_{g} F(q, g, \sigma^2)$
    \State $\sigma^2 \gets \argmax_{\sigma^2} F(q, g, \sigma^2)$
\Until {termination criteria is met}
\Return
$q_1, \ldots, q_p, g, \sigma^2$
\end{algorithmic}
\end{algorithm}

% =================================
\subsection{VEB as penalized linear regression (PLR)}
% =================================
Kim \etal~\cite{kim_mrash_2022} also showed that the VEB optimization problem \eqref{eq:veb} can be recast as a penalized linear regression (PLR) problem in which the penalty depends on the prior $g$, which is estimated from the data. The main contribution of our work is to use this
result to develop an alternative, quasi-Newton approach to solving the VEB problem. In this section we summarize the result from \cite{kim_mrash_2022}, which requires us first to introduce
some notation related to the Normal Means model.

\paragraph{Normal Means Model.}
Let $\NormalMeans{z \mid g, s^2}$ denote the following ``normal means" model for an observation $z$ with prior $g$ and variance $s^2$,
\begin{equation}
\begin{aligned}
    z \mid \mu, s &\sim \mathcal{N}\prth{\mu, s^2}, \\
    \mu \mid g &\sim g \in \G
\label{eq:NM-model}
\end{aligned}
\end{equation}
with $z, \mu \in \realnum$.
Let $\mllNM$ denote the log marginal likelihood of this model,
\begin{equation}
    \mllNM(z, g, s) \triangleq \log \prob{z \mid g, s}
        = \int \log \prob{z \mid \mu, s} g(\mu)\,d\mu\,.
    \label{eq:NM-marginal-log-likelihood}
\end{equation}
Let $\shrinkageop_{g,s} : \realnum \mapsto \realnum$ denote the
posterior mean operator, which maps $z$ to the posterior mean for $\mu$, depending on the prior $g$ and the variance $s^2$:
\begin{align} \label{eq:NM-posterior-mean-shrinkage}
    \shrinkageop[g, s](z) &\triangleq \expectation[{\NM}]{\mu | z, g,s} \\
    &= z + s^2 \mllNM'(z, g, s)\,,
    \label{eq:NM-posterior-mean-shrinkage-tweedie}
\end{align}
where the second line follows from Tweedie's formula~\cite{robbins1956}.
Note that for many prior families $\G$ of interest, both $\mllNM$ and $S_{g,s}$ are analytically tractable \cite{kim_mrash_2022}, which is a key requirement for our methods to work.

\begin{plrtheorem}[Kim et al \cite{kim_mrash_2022}]
Let $\widehat{q}$, $\widehat{g}$, $\widehat{\sigma}^2$ be a solution to
\begin{equation*}
    \widehat{q}, \widehat{g}, \widehat{\sigma}^2 = \argmax_{q \in \mathcal{Q},\, g \in \mathcal{G},\, \sigma^2 \in \realnum_{+}} F(q, g, \sigma^2)
\end{equation*}
and define
\begin{equation*}
    \widehat{\vecb} \triangleq \expectation[\widehat{q}]{\vecb}\,.
\end{equation*}
Then, $\widehat{\vecb}$, $\widehat{g}$, $\widehat{\sigma}^2$ can also be obtained by solving the following optimization problem:
\begin{equation}
    \widehat{\vecb}, \widehat{g}, \widehat{\sigma}^2 = \argmin_{\vbbar \in \realnum^{p},\, g \in \mathcal{G},\, \sigma^2 \in \realnum_{+}} h(\vbbar, g, \sigma^2)\,,
    \label{eq:plr-optimization}
\end{equation}
where the objective function $h(\vbbar, g, \sigma^2)$ has the form of a PLR,
\begin{equation}
\begin{aligned}
    h(\vbbar, g, \sigma^2) &\triangleq - \max_{q:\expectation[q]{\vecb} = \vbbar} F(q, g, \sigma^2) \\
        &= \frac{1}{2\sigma^2} \norm{\vecy - \vecX\vbbar}^2
            + \sum_{j = 1}^{p} \penaltyop[](\bjbar, g, v_j^2)
            - \frac{1}{2} \sum_{j = 1}^{p} \log \prth{d_j^2}
            + \frac{n - p}{2} \log \prth{2 \pi \sigma^2}\,,
\end{aligned}
\label{eq:plr-objective-01}
\end{equation}
in which $d_j^2 \triangleq \prth{\vecx_j^{\transpose}\vecx_j}^{-1}$, $v_j^2 \triangleq \sigma^2 d_j^2$ and the penalty function $\penaltyop[]$ satisfies
\begin{align}
    \penaltyop\prth{\shrinkageop[g, s_j](z_j), g, s_j^2} &= -\mllNM(z_j,  g, s_j) - \frac{1}{2s_j^2}\prth{z_j - \shrinkageop[g, s_j](z_j)}^2
    \label{eq:NM-plr-penalty-Sfs}\\
    \penaltyop'\prth{\shrinkageop[g, s_j](z_j), g, s_j^2} &= \frac{1}{s_j^2}\prth{z_j - \shrinkageop[g, s_j](z_j)}
    \label{eq:NM-plr-penalty-Sfs-deriv}\\
                                                          &= - \mllNM'(z_j, g, s_j)
    \label{eq:NM-plr-penalty-Sfs-deriv-tweedie}
\end{align}
for any $z_j \in \realnum$, any prior distribution $g$ and $s_j \in \realnum_{+}$.
Here, $\mllNM(z_j, g, s_j)$ denotes the marginal log-likelihood of the Normal Means model $\NormalMeans{z_j \mid g, s_j^2}$
defined in (\ref{eq:NM-model}, \ref{eq:NM-marginal-log-likelihood}),
$\shrinkageop[g, s_j](z_j)$ denotes the Posterior Mean operator defined in \eqref{eq:NM-posterior-mean-shrinkage}.
\end{plrtheorem}

% ================================
% =================================
\section{Quasi-Newton Method for Variational Inference}
\label{sec:method}
% =================================
% =================================

Our goal is to solve the VEB optimization problem in \eqref{eq:plr-optimization} using gradient-based methods.
To implement quasi-Newton optimization, one needs to compute the objective function in \eqref{eq:plr-objective-01} and its
derivatives with respect to the optimization parameters, namely $\vbbar$, $g$ and $\sigma^2$.
However, computing the penalty $\penaltyop[](\bjbar, g, v_j^2)$ using the relation \eqref{eq:NM-plr-penalty-Sfs},
requires computing the inverse of the posterior mean operator $S$, which we denote $T$:
\begin{equation}
    \shrinkageopinv(\bjbar, g, v_j^2) \triangleq \shrinkageop[g, v_j]^{-1}(\bjbar).
    \label{eq:inverse-posterior-mean-operator}
\end{equation}
Indeed, combining \eqref{eq:inverse-posterior-mean-operator} and
\eqref{eq:NM-plr-penalty-Sfs}, the penalty function can be written in terms of $T$:
\begin{align}
    \penaltyop[]\prth{\bjbar, g, v_j^2}
        &= \penaltyop[]\prth{\shrinkageop[g, v_j](\shrinkageopinv(\bjbar, g, v_j^2)), g, v_j^2} \\
        &= -\mllNM(\shrinkageopinv[](\bjbar, g, v_j^2), g, v_j) - \frac{1}{2v_j^2}\prth{\shrinkageopinv[](\bjbar, g, v_j^2) - \bjbar}^2\,. \label{eq:penalty_inverse_form}
\end{align}
While $\shrinkageop[g, v_j]$ is available analytically for many prior distributions of interest, its inverse is not generally available (an exception being the normal prior family).
In this work we investigate two different strategies to circumvent this difficulty:
(a) use numerical inversion techniques, and
(b) use a change of variable to make things analytically tractable.

\subsection{Numerical Inversion of Posterior Mean} \label{sec:gradvi-numerical-inversion}
In this approach, we minimize the objective function in \eqref{eq:plr-objective-01}
using numerical approximation to invert the posterior mean operator $S$ and to obtain $T$, see \eqref{eq:inverse-posterior-mean-operator}.
We use the form of the penalty function defined in \eqref{eq:penalty_inverse_form}.
The partial derivatives of $\penaltyop$ are given by:
\begin{align}
    \partial_1 \penaltyop[]\prth{\bjbar, g, v_j^2} &=
        \frac{1}{v_j^2} \prth{\shrinkageopinv[]\prth{\bjbar, g, v_j^2} - \bjbar}\,,
        \label{eq:penalty-objective-deriv-bjbar}\\
    \partial_2\penaltyop[]\prth{\bjbar, g, v_j^2} &=
        - \partial_2 \mllNM\prth{\shrinkageopinv[]\prth{\bjbar, g, v_j^2}, g, v_j^2}\,,
        \label{eq:penalty-objective-deriv-wk} \\
    \partial_3 \penaltyop[]\prth{\bjbar, g, v_j^2} &=
        - \partial_3 \mllNM\prth{\shrinkageopinv[]\prth{\bjbar, g, v_j^2}, g, v_j^2} + \frac{1}{2}\prth{\mllNM'(\shrinkageopinv[](\bjbar, g, v_j^2),g,v_j^2)}^2.
        \label{eq:penalty-objective-deriv-vj2}
\end{align}
Here, \eqref{eq:penalty-objective-deriv-bjbar} follows directly from \equn~\eqref{eq:NM-plr-penalty-Sfs-deriv}.
To derive \eqref{eq:penalty-objective-deriv-wk}, we note that,
\begin{align}
    &\partial_2 \penaltyop[]\prth{\bjbar, g, v_j^2} \nonumber\\
    &\quad=
        \Bigg[
        - \partial_1 \mllNM(\shrinkageopinv[](\bjbar, g, v_j^2), g, v_j^2)\,\partial_2 \shrinkageopinv[](\bjbar, g, v_j^2)
        - \partial_2 \mllNM(\shrinkageopinv[](\bjbar, g, v_j^2), g, v_j^2)
        - \frac{\prth{\shrinkageopinv[](\bjbar, g, v_j^2) - \bjbar}}{v_j^2} \, \partial_2 \shrinkageopinv[](\bjbar, g, v_j^2)
        \Bigg] \nonumber\\
    &\quad= \Bigg[
        - \cancel{\mllNM'(\shrinkageopinv[](\bjbar, g, v_j^2), g, v_j^2) \,\partial_2 \shrinkageopinv(\bjbar, g, v_j^2)}
        - \partial_2 \mllNM(\shrinkageopinv(\bjbar, g), g, v_j^2)
        + \cancel{\mllNM'(\shrinkageopinv[](\bjbar, g, v_j^2), g, v_j^2)\,\partial_2 \shrinkageopinv(\bjbar, g, v_j^2)}
        \Bigg] \nonumber.
\end{align}
In the last step, we used \eqref{eq:NM-posterior-mean-shrinkage-tweedie} to substitute
\begin{equation}
   (\shrinkageopinv[](\bjbar, g, v_j^2)-\bjbar)/v_j^2 = \mllNM'(\shrinkageopinv[](\bjbar, g, v_j^2),g,v_j^2)\,.
    \nonumber
\end{equation}
Finally, \eqref{eq:penalty-objective-deriv-vj2} can be derived using the same strategy as used for \eqref{eq:penalty-objective-deriv-wk}.
Therefore, it becomes possible to calculate the penalty function and its derivatives
with respect to the optimization parameters if we have a numerical way to compute $\shrinkageopinv[](\bjbar, g, v_j^2)$.
The other terms in the objective function \eqref{eq:plr-objective-01} are straightforward
and we can solve the optimzation problem in \eqref{eq:plr-optimization}
using quasi-Newton updates, until a minimum is found.

Several numerical methods exist to compute the inverse of a function, $\shrinkageop[g, v_j]^{-1}\prth{\bjbar}$.
The general idea is to find the root (vector) $\vecz$ of the vector-valued equation
$\shrinkageop[g, \vecv]\prth{\vecz} - \vbbar = \mathbf{0}$, for any given $\vbbar$, $g$ and $\vecv$.
Newton iteration methods, although computationally fast, are not guaranteed to converge.
Bisection methods are guaranteed to converge but computationally expensive.
In our algorithm, we used the trisection method \cite{shammas_trisection}, which reduces the number of iterations compared to the bisection method.
Instead of finding $\shrinkageop[g, v_j]^{-1}\prth{\bjbar}$ for each $j$ sequentially,
we further improved the efficiency by implementing the trisection algorithm in a way such that
$\shrinkageop[g, v_j]^{-1}\prth{\bjbar}$ is updated for all $j$ simultaneously during each update.
For a constant $v_j$ ($v_j = v$ for all $j$), the inversion becomes significantly simpler because
the function $\shrinkageop[g, v]\prth{\cdot}$ is symmetric and monotonic --
allowing us to leverage the fast technique of switching variables
followed by spline interpolation (FSSI)~\cite{tommasini_2020_fssi}.

\subsection{Compound Penalty}
\label{sec:gradvi-compoound-penalty}
In this approach, we reparameterize the optimization problem in terms of variables $z_j = \shrinkageopinv(\bjbar, g, v_j^2)$ (so $\theta_j =\shrinkageop[g, v_j](z_j)$, which we write in vector form as $\vbbar = \shrinkageop[g, \vecv](\vecz)$). The objective function, in terms of $\vecz$, becomes
\begin{equation}
\begin{aligned}
    \htilde(\vecz, g, \sigma^2) &\triangleq h(\shrinkageop[g, \vecv](\vecz), g, \sigma^2) \\
        &= \frac{1}{2\sigma^2} \norm{\vecy - \vecX\shrinkageop[g, \vecv](\vecz)}^2
            + \sum_{j = 1}^{p} \penaltyop[](\shrinkageop[g, v_j](z_j), g, v_j^2)
            - \frac{1}{2} \sum_{j = 1}^{p} \log \prth{d_j^2}
            + \frac{n - p}{2} \log \prth{2 \pi \sigma^2},
\end{aligned}
\label{eq:plr-objective-02}
\end{equation}
and we solve the optimization problem
\begin{equation}
    \widehat{\vecz}, \widehat{g}, \widehat{\sigma}^2 = \argmin_{\vecz \in \realnum^{p},\, g \in \mathcal{G},\, \sigma^2 \in \realnum_{+}} \htilde(\vecz, g, \sigma^2)\,.
    \label{eq:plr-optimization-compound}
\end{equation}
The reparametrization yields an analytically tractable objective function for many priors, because the ``compound penalty function" \eqref{eq:NM-plr-penalty-Sfs} is analytically tractible for many prior families. Once $\widehat{\vecz}$
is obtained we can obtain the solution to the original problem \eqref{eq:plr-optimization} by
\begin{equation}
    \widehat{\vecb} = \shrinkageop[\widehat{g}, \widehat{\vecv}](\widehat{\vecz}).
\end{equation}
The reparametrization to solve for $\vecz$ instead of $\vbbar$
helps to us avoid the numerical inversion during each gradient descent iteration,
but the initialization of $\vecz$ is less straightforward than $\vbbar$ from the previous approach, as discussed below.

\subsection{Choice of prior}
The methods discussed above is generally applicable for any choice of the prior distribution $g$.
However, the tractability of the objective functions (\ref{eq:plr-objective-01}, \ref{eq:plr-objective-02})
and their gradients depend on the tractability of
 $\mllNM$ and its derivative.
Our software GradVI is designed to obtain the prior family as an input module from the user.
For any prior family, one has to define the number of parameters that needs to be estimated,
a real bound for each of those parameters and a function for the gradient descent step for those parameters.
This allows flexibility to implement suitable priors, as compared to coordinate ascent methods.
As examples, we have implemented the following two priors:
\begin{itemize}
    \item \textbf{Adaptive shrinkage (Ash) prior} uses a scale mixture of normals~\cite{stephens_2016_fdr_new_deal}.
    Specifically
    \begin{align}
        \mathcal{G}\prth{\sigma_1^2, \ldots, \sigma_K^2}
            &\triangleq \brcs{g(\cdot) = \sum_{k=1}^{K}w_k\mathcal{N}\prth{\cdot \mid 0, \sigma_k^2} : w_k \in S}\,,
        \label{eq:mrash_prior_g} \\
        S &\triangleq \brcs{w_k \in \realnum_{+}^{K} : \sum_{k=1}^{K}w_k = 1}\,.
        \label{eq:mrash_prior_w}
    \end{align}
    Here, $0 \le \sigma_1^2 \le \ldots \le \sigma_K^2 < \infty$ is a
    pre-specified grid of component variances, and $w_1, \ldots, w_K$ are
    unknown mixture proportions, which are estimated from the data.
    Typically, the first variance $\sigma_1^2$ is set exactly to zero
    to allow for a sparse regression model. In our numerical experiments we followed \cite{kim_mrash_2022} in using $K=20$ and $\sigma_k^2 = (2^{{(k-1)}/K} - 1)^2$.
    \item \textbf{Point-normal prior}, also known as ``spike-and-slab'' prior,
    which is used in ~\cite{carbonetto2012-bvsr} for example.
    It is defined as,
    \begin{equation}
    g\prth{\cdot} = \prth{1 - w}\delta_0 + w \mathcal{N}\prth{\cdot \mid 0, \sigma_1^2}\,,
    \end{equation}
    where $\delta_0$ is the Dirac delta function with a point mass at zero.
    Here, $w$ and $\sigma_1$ are unknown parameters, which are estimated from the data.
\end{itemize}

\subsection{Initialization}

Since the VEB optimization problem is non-convex, the solution obtained may depend on the initialization.
Both approaches (numerical inversion and compound approach) can be initialized by specifying initial values $\vbbar_{\textrm{init}}, g_{\textrm{init}},\sigma^2_{\textrm{init}}$. These can be used to compute initial values $\vecv^2_{\textrm{init}} = \sigma^2_{\textrm{init}} \vecd$ following the definition of $v_j$ in \eqref{eq:plr-objective-01}.
For the compound approach one can then initialize $\vecz$ by computing $\vecz_{\textrm{init}} = \shrinkageop[g_{\textrm{init}}, \vecv_{\textrm{init}}]^{-1}(\vbbar_{\textrm{init}})$ using numerical inversion.

Initial values $\vbbar_{\textrm{init}}$ can be specified either by simply setting $\vbbar_{\textrm{init}}=0$ or by using the estimated regression coefficients from another approach such as lasso. (As shown below in our numerical experiments, the null initialization is good enough
in many use cases, particularly for sparse problems.)
In results presented here we used $\sigma^2_{\textrm{init}}=1$ and initialized mixture priors $g_{\textrm{init}}$ by setting all components to have equal mixture proportions. For example, in the ash prior above we initialize $w_k = 1 / K$ for all $k$.

In cases where $\vbbar_{\textrm{init}}$ is carefully chosen (\eg from lasso) then it may be helpful to perform some initial iterations (of the numerical inversion approach) that update $g$ with $\vbbar,\sigma^2$ held fixed at their initial values. These are relatively low-dimensional updates that can be applied
to quickly improve the initial estimates of $g$.
This strategy is an option in our software, and was used in the numerical experiments below.

\subsection{Software implementation}
Our software GradVI implements the two strategies described above for solving the optimization problem.  We used the L-BFGS-B algorithm~\cite{liu_nocedal_1989_lbfgs, byrd_1995_lbfgsb, lbfgsb_fortran},
a popular quasi-Newton method for minimizing the two objective functions
defined in \eqref{eq:plr-objective-02} and \eqref{eq:plr-objective-01} respectively.
For large $p$ problems the most expensive operation for the function and the derivative evaluations are matrix-vector multiplications of the form $\vecX\vecr$ and $\vecX\transpose\vecr$ for an arbitrary vector $\vecr$. Consequently, if $\vecX$ has special structure that can be exploited to speed up these products then our algorithms can easily exploit this.
Our software is open-source and released under the MIT license.

% ##########################################################
% \input{main/experiments.tex}
\section{Experiments}
\label{sec:experiments}
% ##########################################################

We conducted numerical experiments to compare the convergence and predictive accuracy of our proposed GradVI algorithms with the CAVI algorithm from \cite{kim_mrash_2022}, which is implemented in the R package {\tt mr.ash.alpha}. We follow \cite{kim_mrash_2022} in using the normal mixture (``adaptive shrinkage", or ash) prior \eqref{eq:mrash_prior_g} for these experiments, since they showed that this prior achieves good predictive accuracy in a wide range of settings.

We consider two sets of experiments: the first involves generic high-dimensional ($p>n$) sparse multiple linear regression (\eqref{eq:linear-model}, \eqref{eq:prior-general}), with generic design matrices $\vecX$. In this setting we do not expect GradVI to have a particular advantage over CAVI, but we want to check that it converges to similar optimum and provide similar accuracy as CAVI. The second set of experiments (Section \ref{sec:trend-filter}) involves a Bayesian version of trend filtering \cite{kim_2009_siamrev_l1_trend_filtering}, which can be framed as a high-dimensional sparse multiple regression with a very special $\vecX$ that permits fast matrix-vector products. Since GradVI can exploit this fast matrix-vector product, but CAVI cannot, in this setting we expect GradVI to have a substantial advantage over CAVI.

\subsection{High-dimensional Multiple Linear Regression}
\label{sec:expt-high-dimensional-multiple-linear-regression}

\begin{figure}[!t]
  \begin{center}
    \includegraphics[width=1.0\textwidth]{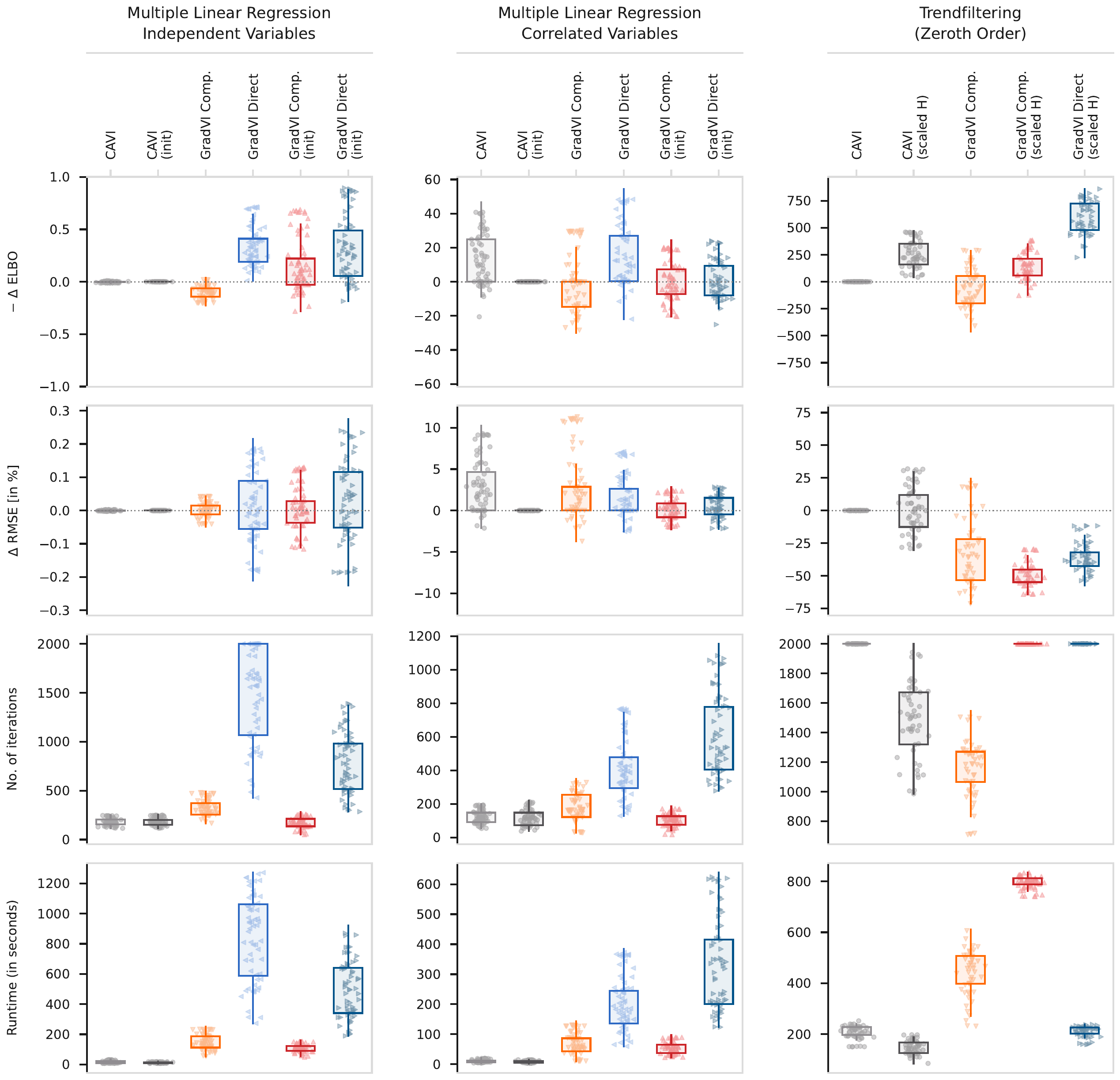}
  \end{center}
  \caption{
    \textbf{Comparison of GradVI and CAVI in different settings.}
    Experiments include high-dimensional multiple linear regression with i.i.d. variables (left column) and
    correlated variables with block-diagonal covariance matrix (center column),
    as well as zeroth order ($k=0$) trendfiltering (right column).
    Evaluation criteria include (rows from top to bottom):
    the absolute difference of optimized ELBO of each method from that of the refence method,
    relative difference of scaled RMSE of predicted responses (on unseen data)
    by the different methods from that predicted by the reference method,
    the number of iterations up to convergence by each method
    and the total time required by each method.
    For multiple linear regression, we use CAVI initialized from lasso as the reference method,
    and for trendfiltering, we use CAVI initialized from genlasso as the reference method.
    All methods were allowed a maximum of 2000 iterations.
    Description of methods: ``CAVI'' uses \code{mr.ash.alpha} with zero initialization.
    ``GradVI Comp.'' uses our method with compound penalty
    and ``GradVI Direct'' uses our method with direct penalty.
    The descriptor ``init'' refers to intialization of the methods -- we used lasso for initializing multiple linear regression
    and genlasso for initializing trendfiltering.
    The descriptor ``scaled H'' refers to using a scaled predictor matrix for the trendfiltering problem.
    }
  \label{fig:linreg-indep-elbo-rmse}
\end{figure}

For this initial set of experiments we generated a design matrix $\vecX \in \realnum^{n \times p}$ with $n = 500$ samples and $p = \mbox{10,000}$ predictors.
We considered two types of design matrix:
\begin{itemize}
  \item \textit{Independent variables}, in which individual observations $x_{ij}$ were simulated \textit{i.i.d.}
  from the standard normal distribution.
  \item \textit{Block-correlated variables}, in which each row of $\vecX$ was an independent draw from a multivariate normal distribution
  with mean zero and covariance matrix $\vecSigma$.
  The covariance matrix $\vecSigma$ was block diagonal, split into 3 blocks of random size,
  such that the minimum size of each block is 2,000 predictors.
  The off-diagonal elements in each block of $\vecSigma$ was chosen to be 0.95 and the diagonal of $\vecSigma$ was set to 1.
\end{itemize}
We then selected $s$ causal predictors uniformly at random from the $p$ variables.
We sampled non-zero coefficients $b_j$ for the selected $s$ causal predictors from the standard normal distributions,
while the coefficients for all other predictors were assumed to be zero.
Finally, we simulated the responses according to \eqref{eq:linear-model}.
For each simulation, the noise variance $\sigma^2$ was set to
$\sigma^2 = \mathrm{Var}(\vecX\vecb) \times (1 - \mathrm{PVE})/{\mathrm{PVE}}$ to hit the target $\mathrm{PVE}$.
In our simulations, we varied the number of causal predictors, $s = \brcs{2, 5, 10, 20}$
and the proportion of variance explained, PVE $ = \brcs{0.4, 0.6, 0.8}$.

We performed VEB regression with ash prior on all simulation replicates,
using CAVI and the two variants of GradVI: (a) ``GradVI Compound'' (using reparametrized objective function, \secn\ref{sec:gradvi-compoound-penalty}), and
(b) ``GradVI Direct'' (using numerical inversion of the penalty function, \secn\ref{sec:gradvi-numerical-inversion}).
We used the R software \code{mr.ash.alpha} for the CAVI algorithm.
For the ash prior, we used the settings recommended in \cite{kim_mrash_2022}.
In short, we chose $K=20$ pre-specified components with variances $\sigma_k^2 = (2^{(k-1)}/20 - 1)^2$.
For all the three methods, we compared the effect of initializing the coefficients $\vecb$
to the Lasso solution $\vecb_{\textrm{lasso}}$ after choosing the Lasso penalty via cross-validation,
because this setting provided better performance than the null initialization
($\vecb_{\textrm{init}} = \mathbf{0}$) in CAVI.
For evaluating optimization performance, we chose two criteria:
(a) the absolute difference of ELBO obtained after optimization using each method from that of the reference method:
\begin{equation}
  \Delta\,\textrm{ELBO} = \textrm{ELBO}_{\textrm{method}} - \textrm{ELBO}_{\textrm{ref}}\,,
\end{equation}
and (b) the relative difference of scaled root mean squared error (RMSE) for predicting responses ($\widehat{\vecy}_{\textrm{test}}$) in unseen test data ($\vecX_{\textrm{test}}$):
\begin{align}
  \textrm{RMSE} \prth{\vecy_{\textrm{test}}, \widehat{\vecy}_{\textrm{test}}} &= \norm{\vecy_{\textrm{test}} - \widehat{\vecy}_{\textrm{test}}} / \sqrt{n} \, \nonumber\\
  \Delta\,\textrm{RMSE} &= \frac{\textrm{RMSE}_{\,\textrm{method}} - \textrm{RMSE}_{\,\textrm{ref}}}{\textrm{RMSE}_{\,\textrm{ref}}} \times 100 \,.
\end{align}
In the above equations, the subscript \code{ref} corresponds to the reference method,
which is CAVI initialized from the Lasso solution. Since this reference method outperformed many other approaches to sparse regression in \cite{kim_mrash_2022}, performing as well as this reference reflects competitive performance.
For evaluating computational performance, we looked at the total number of iterations
and the total runtime in seconds.

We show the results of the experiments in the left and middle columns of \figr~\ref{fig:linreg-indep-elbo-rmse}.
Each row corresponds to a different measure of performance.
For independent variables, all the methods had very similar ELBO and RMSE.
GradVI Compound with null initialization provided slightly better ELBO than the other methods.
GradVI Direct required more iterations than the other methods and
we observed a few simulations where the quasi-Newton algorithm did not converge after 2000 iterations.
For the GradVI methods, null initialization provided slightly better ELBO
and had less variance in RMSE than the Lasso initialization,
although the latter required fewer iterations to converge.

For correlated variables, GradVI compound generally converged to a better ELBO, and  in fewer iterations, than GradVI Direct; the two had similar performance in terms of RMSE.
However, GradVI Compound sometimes failed to converge after 2000 iterations,
whereas GradVI Direct converged in all simulations.
In contrast to CAVI, the null initialization for GradVI had better ELBO than the Lasso initialization.
In general, all GradVI methods performed better than the CAVI null initialization
probably because GradVI updates all parameters at once.

\subsection{Bayesian Trend Filtering} \label{sec:trend-filter}

Trend filtering is a method for finding an estimate $\widehat{\vecmu}$ of $f(\vecx)$ in the
following non-parametric regression model:
\begin{align}
  \label{eq:nonparametric-model}
  y_i &= f\prth{x_i} + \epsilon_i, \quad i = 1, \ldots, n, \\
  \epsilon_i &\sim \mathcal{N}\prth{\,\cdot \mid 0, \sigma^2}  \nonumber
\end{align}
where $x_1, \ldots, x_n \in \realnum^n$ are evenly spaced inputs over the interval $\sqbr{0, 1}$,
and $f: \sqbr{0, 1} \mapsto \realnum$ is the underlying function to be estimated.

Specifically, $\ell_1$ trend filtering,
which was introduced by Kim \etal~\cite{kim_2009_siamrev_l1_trend_filtering}, is shown by \cite{tibshirani_2014_annstat_trend_filtering}
to be equivalent to the following Lasso problem:
\begin{equation}
  \label{eq:l1-trendfiltering-plr-form-synthesis}
  \widehat{\vecb} = \argmin_{\vecb \in \realnum^n} \frac{1}{2}\norm{\vecy - \vecH^{(k + 1)}\vecb}_2^2 + \lambda \sum_{j=k+2}^{n}\abs{b_j}\,,
\end{equation}
 where the matrix $\vecH^{(k+1)}$ depends on the order $k$ of the trend filtering.
For $k=0$,
\begin{equation}
\vecH^{(1)} =
\begin{bmatrix}
  1 & 0 & 0 & \ldots & 0 \\
  1 & 1 & 0 & \ldots & 0 \\
  1 & 1 & 1 & \ldots & 0 \\
  \vdots & & & \ddots & \vdots \\
  1 & 1 & 1 & \ldots & 1
\end{bmatrix}.
\label{eq:trendfiltering-H1-matrix}
\end{equation}
In this special case trend filtering corresponds to assuming that $f$ is piecewise constant, and the non-zero elements of $\vecb$ determine the locations at which the function $f$ changes values (``changepoints"). For general $k$ the form of $\vecH^{(k)}$ is given in \cite{tibshirani_2014_annstat_trend_filtering}.
The matrix $\vecH^{(1)}$ has special structure that allows $\vecH^{(1)}\vecv$ to be computed efficiently\footnote{eg in R or numpy $\vecH^{(1)} v$ is {\tt cumsum}($\vecv$)}, in $O(n)$ operations. More generally, for moderate $k$ $\vecH^{(k)}$  also allows fast matrix-vector multiplication ($\vecH^{(k)}\vecv$ involves $O(kn+2(k - 1)^2)$ operations in our implementation).

To summarize the above: $k$th order ($\ell_1$) trend-filtering is equivalent to fitting the multiple regression model \eqref{eq:linear-model} with an $\ell_1$ penalty on $\vecb$ and a predictor matrix $\vecX=\vecH^{(k+1)}$ that admits fast matrix vector products.

\begin{figure}[!t]
  \includegraphics[width=\textwidth]{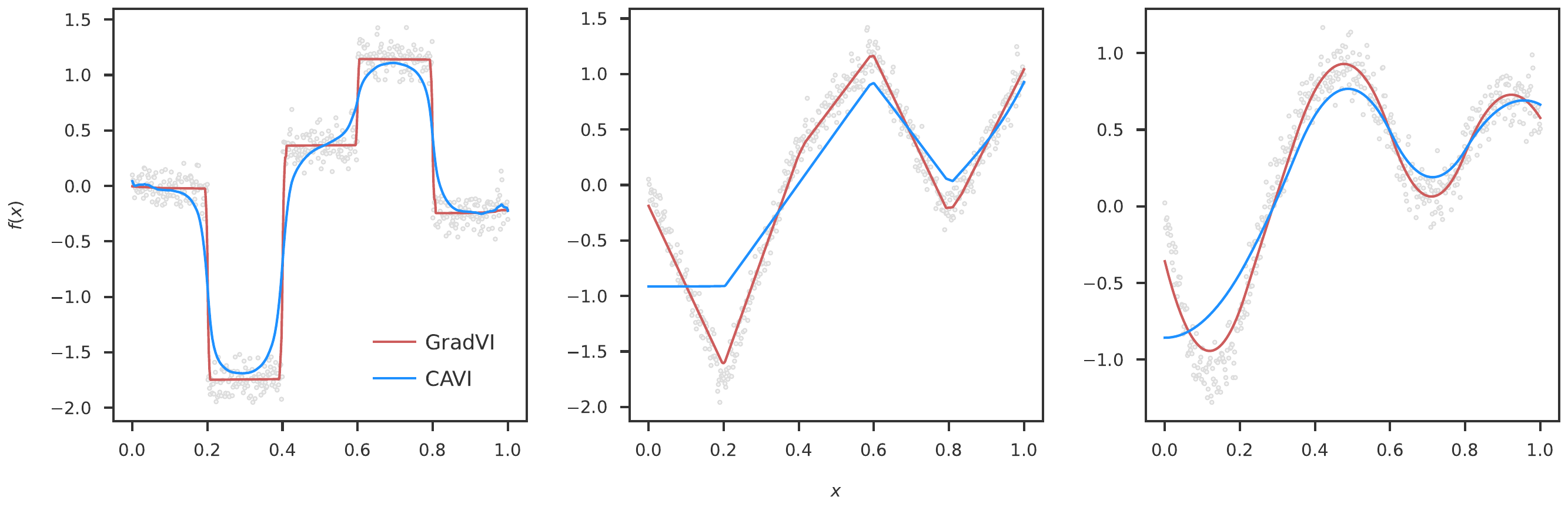}
  \caption{
    \textbf{Simple examples of trend filtering using GradVI.} From left to right, we show trend filtering for constant, linear and quadratic orders ($k=0,1,2$).
    The observed data is shown in grey dots, the signal estimated by GradVI is shown by the red solid line and the signal estimated by CAVI is shown by the blue solid line. The improved fit of GradVI is visually evident in all three cases. Since both methods are optimizing the same objective function, this improved fit reflects the superior convergence of GradVI vs CAVI in these problems.
    }
  \label{fig:trendfiltering-simple-examples-gradvi}
\end{figure}

Here we take the natural step of defining an EB version of $k$th order trend-filtering, by fitting the EB multiple regression model \eqref{eq:linear-model}-\eqref{eq:prior-general} with predictor matrix $\vecX=\vecH^{(k+1)}$.
This EB version of trend filtering has features that make our gradient-based GradVI approach far preferable to the coordinate-based CAVI algorithm used in \cite{kim_mrash_2022}. First, and most important,
 the matrix-vector products $\vecH \vecb$ and $\vecH \transpose \vecb$ can be solved very efficiently, with linear complexity in the number of data points $n$, and without explicitly forming $\vecH$. The CAVI algorithm cannot exploit this fact, but GradVI can. Thus CAVI has quadratic complexity and memory requirements in $n$ whereas GradVI has linear complexity in $n$. Second, the columns of $\vecH$ are very highly colinear, a situation that makes coordinate-based optimization particularly challenging (see Figure \ref{fig:2d-optim-steps}). As a result, trend filtering fits obtained from GradVI are often qualitatively better than those obtained from CAVI; see \figr~\ref{fig:trendfiltering-simple-examples-gradvi} for some simple illustrative examples.

We performed numerical experiments to illustrate the advantages of GradVI over CAVI for this problem.
For each experiment we generated data from the non-parametric regression model  \eqref{eq:nonparametric-model} with $f$ assumed to be piece-wise constant (ie zeroth-order trendfiltering).
We set the number of inputs $n = 4096$, the number of changepoints in $f$ to be $10$ and varied 3 parameters:
\begin{itemize}
  \item The locations of the changepoints, which were chosen uniformly at random.
  \item The magnitude of the changes in $f$ at the changepoints, which were sampled from a Gaussian distribution with mean $0$ and variance $1$.
  \item The noise parameter $\sigma$, which was chosen uniformly from $(0.2, 0.6, 1.0, 1.4, 1.8)$. We repeated the experiment $25$ times for each noise parameter $\sigma$.
  Note that this will give different signal-to-noise ratio for each experiment because the signal is also varying.
\end{itemize}

\begin{figure}[!t]
  \includegraphics[width=\textwidth]{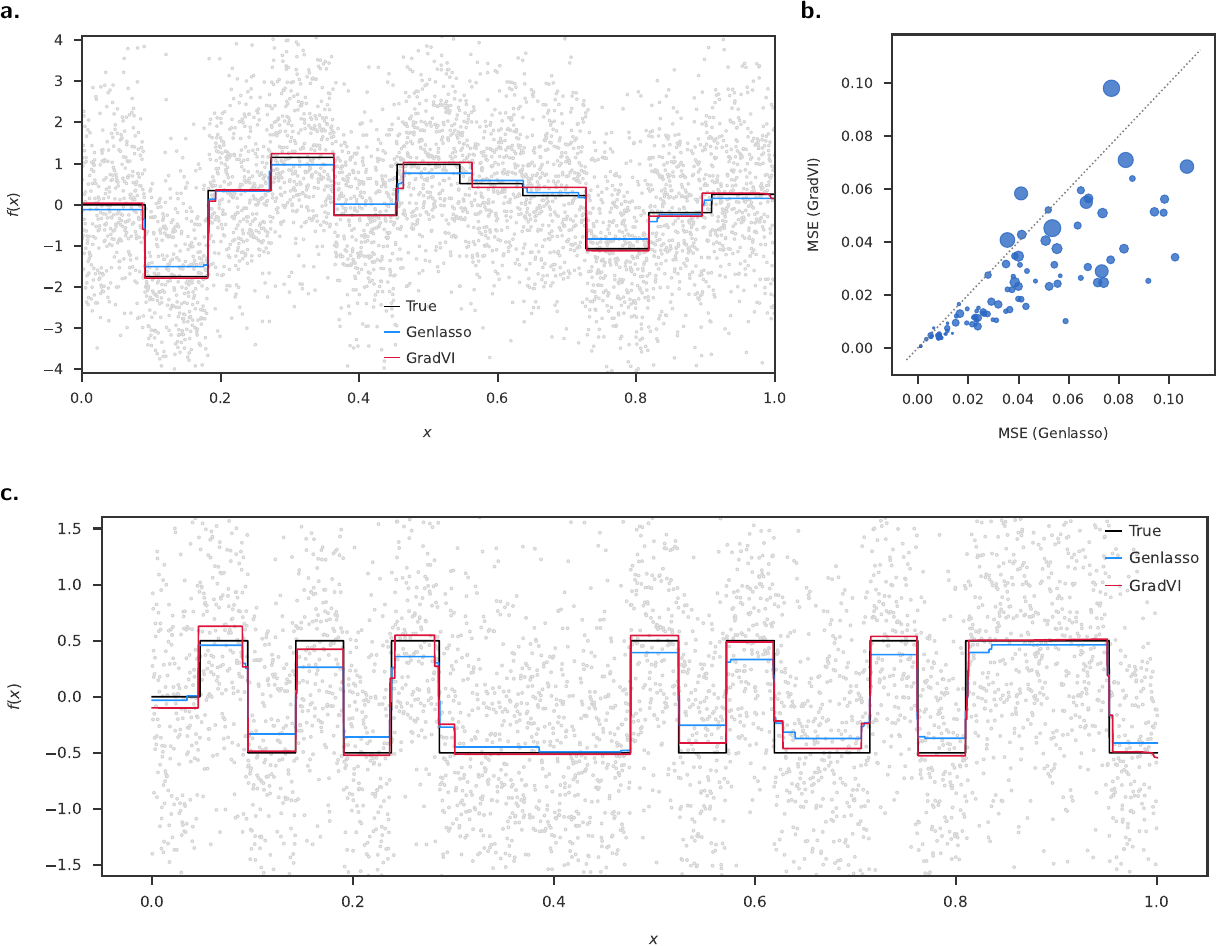}
  \caption{
    \textbf{Comparison of GradVI and genlasso for zeroth order trend filtering.}
    Results from our experiments for zeroth order trend filtering with $n=4096$ and $10$ changepoints,
    with different location and magnitude of the changepoints and different signal-to-noise ratio.
    (a) One representative example of the underlying true block function used for the simulations,
    and the estimated trend using GradVI and genlasso.
    (b) Comparison of the mean square error (MSE) between the true function and the trend estimated by genlasso (x-axis)
    and GradVI (y-axis). Each circle is one simulation, and the area of each circle is inversely proportional to the signal-to-noise ratio.
    (c) One representative example designed with equal magnitude of the changepoints to show
    how the genlasso method systematically underestimates the magnitude of the changepoints.
    }
  \label{fig:trendfiltering-accuracy-genlasso-gradvi}
\end{figure}

First we compared GradVI with CAVI; see right column of \figr~\ref{fig:linreg-indep-elbo-rmse}.
For performance comparison, we used the same metrics as in the multiple linear regression comparisons above,
but we computed RMSE from the ``true'' underlying trend without noise
instead of a test data with noise.
For the design matrix, we used the $\vecH^{(1)}$ matrix  $\eqref{eq:trendfiltering-H1-matrix}$
as well as a scaled version of the $\vecH^{(1)}$ matrix such that all columns of $\vecH^{(1)}$
has the same variance, that is, $\vech_j\transpose\vech_j = d$ for all $j$.
Using a scaled $\vecH^{(1)}$ matrix deviates slightly from the usual trendfiltering approach, but allows fast inversion of the posterior mean operator
(See Section \ref{sec:gradvi-numerical-inversion})
and hence substantially reduces the runtime of the GradVI Direct
method. (GradVI Direct is otherwise computationally intensive and so we did not run it here.)
All methods were initialized from the solution obtained after running \code{genlasso}.

For all simulations, the CAVI algorithm did not converge after 2000 iterations
when using the $\vecH^{(1)}$ matrix, but it converged when using the scaled version. However,
the converged ELBO from CAVI with the scaled $\vecH^{(1)}$ matrix
was lower (worse) than those obtained after 2000 iterations using the $\vecH^{(1)}$ matrix, so we used CAVI with the unscaled $\vecH^{(1)}$ matrix as the ``reference'' method
for the ELBO and RMSE metrics.
The GradVI algorithms with the scaled $\vecH^{(1)}$ matrix did not converge after 2000 iterations,
but the GradVI Compound converged to the best ELBO when using the $\vecH^{(1)}$ matrix.
The RMSE obtained from all the GradVI methods were significantly better than the CAVI methods.

\begin{figure}[!t]
  \centering
  \includegraphics[width=0.5\textwidth]{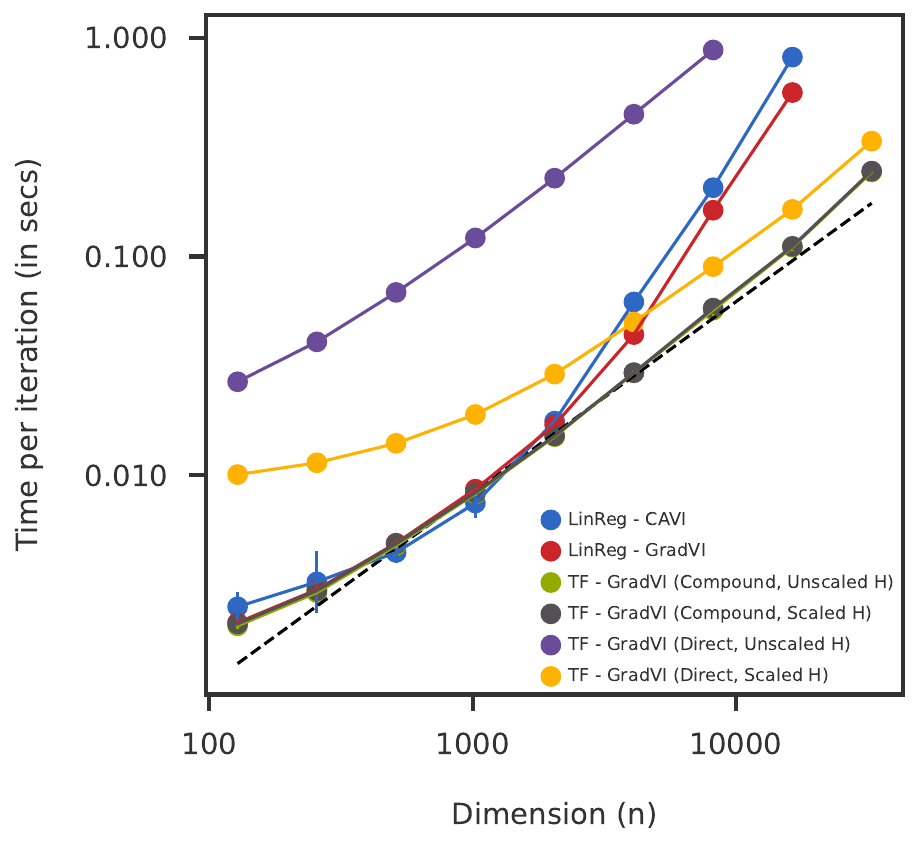}
  \caption{
    \textbf{Speed comparison of Bayesian trend filtering methods.}
    We show the dependence of per-iteration time (in seconds) on the dimension ($n$)  of the trend filtering problem
    for GradVI compound, GradVI direct and CAVI.
    For each method, we also compare the dependence on scaled and unscaled $\vecH$ matrix for trend filtering.
    Both axes are on a $\log$ scale.
    Faster matrix-vector multiplication improves per-iteration speed of GradVI compound for both scaled and unscaled $\vecH$.
    In GradVI direct, the per-iteration speed increases linearly with the dimension but there is a substantial overhead for inversion of the posterior mean operators when using unscaled $\vecH$.
    }
  \label{fig:trendfiltering-per-iteration-time}
\end{figure}

To illustrate the gains that come from exploiting the fast matrix-vector products \figr~\ref{fig:trendfiltering-per-iteration-time}, compares the speed per iteration of GradVI
using two different implementations:
(a) \textit{LinReg - GradVI} uses the standard matrix-vector multiplication, as implemented in \code{numpy},
and (b) \textit{TF - GradVI} uses the faster implementation of the matrix-vector multiplication.
As noted above, CAVI cannot exploit the fast matrix vector multiplication, so we show the standard implementation, denoted by \textit{LinReg - CAVI}.
We compare all variants of our method using the compound penalty and the direct penalty,
as well as scaled and unscaled $\vecH$ matrix.

As expected, faster matrix-vector multiplication improves the speed per iteration of all variants of GradVI. For the direct method with unscaled $\vecH$ there is a high cost per iteration
because of the slow numerical inversion.
For the direct method with scaled $\vecH$, the inversion is fast (using our FSSI implementation)
and can achieve faster speed per iteration.
For the compound method, both scaled and unscaled $\vecH$ have similar speed because there is no numerical inversion involved.

Although our major focus here is to compare GradVI with CAVI, it is natural to wonder about the performance of our EB approach to trend filtering vs the original $\ell_1$ trend filtering approach based on Lasso penalty. We therefore also
compared GradVI with $\ell_1$ trend filtering implemented in R package \code{genlasso},
using cross validation to tune the $\ell_1$ penalty.
In \figr~\ref{fig:trendfiltering-accuracy-genlasso-gradvi}a,
we show a representative example of the underlying block functions used for the simulation.
The estimates obtained using genlasso and GradVI Compound are also shown in the plot.
Both GradVI and genlasso can succesfully detect most of the changepoints despite the noisy data.
In \figr~\ref{fig:trendfiltering-accuracy-genlasso-gradvi}b, we compare the accuracy of the two methods
using the mean square error (MSE) with respect to the true function.
We found that GradVI obtained better accuracy (lower MSE) than genlasso,
over a wide range of noise (size of each point).
The performance difference become more prominent with increasing noise.
We attribute this improvement of performance to the better estimation of the magnitude of the changepoints.
In \figr~\ref{fig:trendfiltering-accuracy-genlasso-gradvi}c, we designed an experiment with equal magnitudes of the changepoints
to show that the lasso method underestimates the magnitudes systematically.

% ##########################################################
% \input{main/discussion.tex}
\section{Discussion}
\label{sec:discussion}
% ##########################################################
In this study, we have presented a novel gradient-based optimization method
for variational inference (GradVI) in multiple linear regression.
GradVI complements the conventional coordinate ascent variational inference (CAVI), and can show superior performance in some settings, including when the variables are highly correlated,
or when the design matrix yields fast matrix-vector multiplications,
\eg trend filtering.

The main challenge in implementing our gradient-based approach is that the penalty function is not analytically tractible. We developed two variants of GradVI, Compound and Direct, which address this problem in different ways.
The optimal choice between these variants depends on the specific problem at hand.
Generally, we recommend using the compound penalty,
which is the default setting in our software,
due to the numerical challenges associated with the functional inverse required in the direct method.
However, in cases where the matrix of explanatory variables is well-conditioned and uniform,
the direct method may be more effective
(e.g.~we have seen this in higher order, $k \geq 1$, trendfiltering problems,
which we have not extensively covered in this manuscript).
For situations where users are constrained to null initialization,
such as when cross-validation is computationally prohibitive,
GradVI may be preferable to standard CAVI.

The recent trend towards utilizing gradient-based optimization techniques
in variational inference aims to achieve efficient and scalable posterior inference
in advanced Bayesian computation.
While low-dimensional illustrations often motivate the choice of approximating family and divergence,
these intuitions do not necessarily extend to high-dimensional settings.
For example, stochastic variational inference~\cite{hoffman_2013_jmlr_svi}
assumes that the prior distribution belongs to the exponential family,
which is suboptimal for high-dimensional scenarios
that require efficient sparsity-inducing priors (\eg slab-and-spike prior, adaptive shrinkage prior).
Methods like black-box variational inference~\cite{ranganath_2014_pmlr_black_box_vi}
and automatic differentiation variational inference~\cite{advi_kucukelbir}
sometimes fail to converge due to the high variance of nested gradient estimates.
Moreover, optimizing the KL divergence in these methods necessitates updating all variational parameters,
which becomes unmanageable as dimensionality increases.

GradVI addresses this issue by optimizing the mean of the approximate variational posterior,
thereby reducing the number of parameters to be optimized and enhancing computational scalability.
For example, GradVI enables trend filtering for datasets with $n=10^6$
on a single core with 100 GB of memory, a task where both genlasso cross-validation
and CAVI methods exhaust available memory.
Our software leverages the L-BFGS-B method for numerical optimization,
though Newton’s method could further accelerate computations
if the Hessian for the penalty function is implemented.
While we have performed some initial experiments that confirm the potential speed benefits of Newton's method,
it is not yet integrated into our software.

Another significant challenge in variational inference methods
for sparse multiple linear regression is their sensitivity to initialization,
particularly with correlated variables.
Therefore, in most cases, variational inference methods
are initialized from Lasso results.
In large datasets, where it becomes computationally challenging to run Lasso due to the cross-validation,
GradVI offers a notable advantage over CAVI in terms of accuracy,
as it is less sensitive to initialization than CAVI.

A significant advantage of GradVI is its simplicity and flexibility in handling different prior distributions.
The CAVI updates for different prior distributions must be derived and implemented separately,
which involves substantial effort in terms of derivation, programming, debugging, and scaling.
This complexity often prevents full utilization of the benefits associated
with exploring different prior distributions for any regression problem.
In contrast, implementing any prior distribution with GradVI requires solving
a relatively simpler Normal Means problem, making it a more practical choice for practitioners.

GradVI is a robust and scalable alternative for high-dimensional linear regression
and Bayesian trend filtering.
The promise of GradVI extends beyond the scope of this study.
Future work will focus on integrating Newton's method for further speed improvements
and exploring its application in other complex Bayesian models.
Additionally, we plan to incorporate automatic differentiation tools
to further simplify the implementation of different priors.
We are excited about the potential of GradVI to make advanced Bayesian methods
more accessible and efficient for a wide range of applications.

% ##########################################################
% \input{main/metainfo.tex}
% ##########################################################
\section*{Code Availability}
\label{sec:code-availability}
% =================================
% =================================

GradVI software is available on Github (\url{https://github.com/stephenslab/gradvi}).
The numerical experiments were performed with Dynamical Statistical Comparisons
(DSC, \url{https://github.com/stephenslab/dsc}) for reproducibility.
The DSC analysis codes and scripts are available at
\url{https://github.com/banskt/gradvi-experiments}.
For comparison with CAVI, we used the R package \code{mr.ash.alpha} version 0.1.43
(\url{https://github.com/stephenslab/mr.ash.alpha}).
For $\ell_1$ trend filtering, we used the R package \code{genlasso} version 1.6.1
(\url{https://cran.r-project.org/package=genlasso}).

\section*{Acknowledgments}
\label{sec:acknowledgements}
% =================================
% =================================

We thank Dongyue Xie for helpful discussions.
This work was supported by the NHGRI at the National Institutes of Health under award number R01HG002585 to Matthew Stephens.
Computing resources were provided by the University of Chicago
Research Computing Center.

% ========================================================== 
% Bibliography is journal-specific
% \printbibitembibliography
% ========================================================== 

% \printbibliography

%\nolinenumbers
\end{document}

\typeout{get arXiv to do 4 passes: Label(s) may have changed. Rerun}